\documentclass[10pt]{article}

\usepackage{amsmath,bm}
\usepackage{amsthm}
\usepackage{graphicx,pdflscape}
\usepackage{color}
\usepackage[toc,page]{appendix}
\usepackage{simpler-wick}

\usepackage{color,subfigure}

\usepackage{hyperref}
\hypersetup{
    colorlinks,
    citecolor=red,
    filecolor=cyan,
    linkcolor=blue,
   urlcolor=magenta
 }

\usepackage{xparse}
\ExplSyntaxOn
\NewDocumentCommand{\mref}{m}{\quinn_mref:n {#1}}
\seq_new:N \l_quinn_mref_seq
\cs_new:Npn \quinn_mref:n #1
 {
  \seq_set_split:Nnn \l_quinn_mref_seq { , } { #1 }
  \seq_pop_right:NN \l_quinn_mref_seq \l_tmpa_tl
  ( 
  \seq_map_inline:Nn \l_quinn_mref_seq
    { \ref{##1},\nobreakspace } 
  \exp_args:NV \ref \l_tmpa_tl 
  ) 
 }
\ExplSyntaxOff

\newcommand{\disp}[1]{Eq.~\mref{#1}}

\newcommand{\figdisp}[1]{Fig.~\mref{#1}}

\newcommand{\lessim} {\ {\raise-.5ex\hbox{$\buildrel<\over\sim$}}\ }

\newcommand{\gssim}{\ {\raise-.5ex\hbox{$\buildrel>\over\sim$}}\ }

\newcommand{\tJ}{\ $t$-$J$ \ }
\newcommand{\nn}{\nonumber}

\renewcommand{\Re}{\mathrm{Re}}
\renewcommand{\Im}{\mathrm{Im}}

\renewcommand{\emph}{\textit}

\renewcommand{\k}{\vec{k}}

\usepackage{hyperref}
\usepackage{color}

\newcommand{\change}[1]{{\color{black} #1}}

\newcommand{\beq}{\begin{eqnarray}}
\newcommand{\eeq}{\end{eqnarray}}
\newcommand{\barray}{\begin{eqnarray}}
\newcommand{\earray}{\end{eqnarray}}
\def\Xint#1{\mathchoice
   {\XXint\displaystyle\textstyle{#1}}%
   {\XXint\textstyle\scriptstyle{#1}}%
   {\XXint\scriptstyle\scriptscriptstyle{#1}}%
   {\XXint\scriptscriptstyle\scriptscriptstyle{#1}}%
   \!\int}
\def\XXint#1#2#3{{\setbox0=\hbox{$#1{#2#3}{\int}$}
     \vcenter{\hbox{$#2#3$}}\kern-.5\wd0}}

\def\dashint{\Xint-}
\newcommand{\half}{\frac{1}{2}}
\newcommand{\bw}{\bar{\omega}}
\newcommand{\btau}{\bar{\tau}}
\newcommand{\bA}{\bar{A}}
\newcommand{\wert}{\vert_{T=0}}
\newcommand{\AW}{\Omega_*}
\newcommand{\AAW}{W_0}

\usepackage{graphicx} 

\usepackage{amsthm}
\theoremstyle{plain}
\newtheorem*{theorem*}{Theorem}

\begin{document}


\title{ Method for reconstructing the  self-energy  from the spectral function}

\author{ B Sriram Shastry$^{1}$\footnote{sriram@physics.ucsc.edu}  \\
\small \em $^{1}$Physics Department, University of California, Santa Cruz, CA, 95064 \\}
\date{\today}

\maketitle

\abstract

A fundamental question about the nature of   quantum materials such as High-T$_c$ systems remain open to date- it is unclear  whether they are (some variety of) Fermi liquids, or  (some variety of) non Fermi liquids. A  direct avenue to determine their nature is to study the (imaginary part of the) self-energy  at low energies.  Here we  present a method to extract this  low $\omega$ self-energy from experimentally derived spectral functions. The method seems suited for  implementation with high quality angle resolved photoemission data.  It is based on a helpful {\em Theorem} proposed here, which assures us that the method   has  minimal  (or vanishing) error at the  lowest energies.   We provide numerical examples  showing that  a few popular  model systems yield distinguishably different low energy self-energies.


\section{Introduction}
In this work we address the basic problem of reconstructing  the low energy   $\Im \, \Sigma(\k_F,\omega)$ from the  spectral function $A(\k,\omega)$   inferred from angle resolved photoemission experiments (ARPES). We   refer to this as the   {\em inversion problem} in this work. The ARPES probe of quantum materials \cite{Sobota}  is known to play  a vital part in our understanding of the important class of strongly correlated materials. 
The low-$\omega$ dependence of this object for $\k$$\sim$$k_F$ is of especial interest in theoretical studies, since  reliable high precision measurements, if available,  would provide an essential direction in the search for  a suitable theory for systems with strong correlations, and possibly also for superconducting states.  Physically the low $\omega$ object is directly related to the decay rate of a slightly excited  particle near the ground state, and also can be used to infer  a significant contribution to the $T$ dependence of the resistivity of the system at low $T$. 


\change{An overarching  question posed by numerous experimental results on strongly correlated systems, is whether these can be described using standard  methods of condensed matter  physics, or  if not, whether they demand substantial revisions of the otherwise highly successful standard theory. The latter is  based on the  density functional theory of Kohn {\it et al.} as a starting point, and by treating the effects of interactions using perturbative expansions in the spirit of  Landau's Fermi liquid theory. Its validity can be questioned if the interactions become very strong,  leading to much debate in the community. 
To sharpen the debate one   chooses the most sensitive class of experimental results, and sees whether the predictions of standard theory are violated, and if so to what extent. Much emphasis so far has been on the temperature dependent resistivity, which indeed has surprising and unexpected aspects in many strongly correlated systems. However, the theoretical calculation of the resistivity has considerable intrinsic  complexity, being a higher order correlation function involving {\em two electrons and two holes}. Therefore the evaluation of the (formally exact) Kubo formula for resistivity can only be done approximately, the difficulty being exacerbated   when the interaction scales are large. Additionally  there are system-dependent  details  at play, such as the different $T$ dependences at different densities.  Altogether    these factors  seem to prevent rigorous conclusions from being drawn on the  question we started with, in either direction.

On the other hand, the relatively simpler  {\em single electron and one hole}  spectral function contains crucial information about the low-energy and low-temperature dependence of the lifetime of a particle excited above the ground state. Extracting or accessing the  most relevant pieces of information from the ARPES data is, however, a non-trivial task. This goal  has been achieved at a certain level of accuracy, using an elaborate collation of energy dependence of spectral data at fixed wavevectors as described below. The purpose of this work is to introduce a new  and  direct method for this task, using an exact but rarely used formula for the self-energy in terms of the spectral function \disp{Eq-1,def-Phi}. We also show that this method becomes increasingly accurate -- even asymptotically exact -- at lowest energy.  By applying this method to a quantum material studied in its normal state,  we  could unravel essential details of its low-energy  behavior, and thus
ascertain   whether it is a fermi liquid or a quantum liquid of some other sort, and thereby provide definitive answers to the    initially posed question.

 }

\subsection{Current status of the inversion problem}
At present a few works  report such an inversion.   Important recent examples are given in   Ref. \cite{Valla,Ashoori}. The original effort of Ref. \cite{Valla} uses the direct relationship between the spectral function as a function of the self energy:
\beq
A(\k,\omega)= \frac{-1}{\pi} \frac{\Im \,\Sigma(\k,\omega)}{(\omega+\mu-\varepsilon_k-\Re\, \Sigma(\k,\omega))^2+ (\Im\, \Sigma(\k,\omega))^2},
\eeq
and therefore at a fixed $\omega$ the width of a peak $\Delta k_{w}$ is given by
\beq
\hbar v^{eff}_{k} \Delta k_{w}\sim 2 \Im\, \Sigma(\k,\omega)
\eeq
where $v^{eff}_k$ is a renormalized velocity, so that the left hand side may be roughly  estimated from experiments.
\change{This} inversion problem of reconstructing $\Im \, \Sigma(\k,\omega)$, requires the collation of data from several constant energy sections of the spectra function, termed  the momentum distribution curves (MDC). A second method is presented in Ref. \cite{Ashoori}, which uses a novel momentum-energy resolved tunneling method, and demonstrate its working for two-dimensional electron systems embedded in a semiconductor. This important class of materials  appears to be ideally suited for that method.

\subsection{A  proposal for  inversion \label{proposal}}
The present work discusses  a different framework for effecting this inversion, and provides some examples of how it can be used. In the following instead of the imaginary part of the self-energy $ \Im \, \Sigma(\k,\omega)$,  we discuss
 the equivalent but more convenient positive definite spectral density of the self-energy,  obtained from  $\rho_\Sigma(\k,\omega)= -\frac{1}{\pi} \Im \, \Sigma(\k,\omega)$.
  We begin with  an exact   but apparently infrequently used \cite{AnatomyPaper} relation   
  \beq
\rho_\Sigma(\k,\omega)&=&\frac{ A(\k,\omega)}{\{ \Phi_{\k}(\omega)\}^2 +  \{\pi A(\k,\omega) \}^2} \label{Eq-1}\\
\Phi_{\k}(\omega)& \equiv& \dashint d\nu \frac{A(\k,\nu)}{\omega-\nu} \label{def-Phi}
\eeq
where 
$A(\k,\omega)$ is the ARPES related spectral function,  $\Phi_{\k}(\omega)$ is  the
real part of the Greens function  \disp{self-energy-fromspectral-11}, and  $\dashint$ represents a principal value integration (see Appendix (\ref{app-1}) for details).
We  see from \disp{Eq-1,def-Phi} that $\rho_\Sigma(\k,\omega)$ is a  functional of $A(\k,\nu)$. If the overall scale of $A(\k,\omega)$ is unknown, as is usually the situation in experiments, we can only hope to obtain $\rho_\Sigma$ up to an overall constant. 
 The non-locality (in frequency) of \disp{def-Phi} presents the main obstacle in this route of inversion.  Although $A(\k,\omega)$ is experimentally available  for a range of $\omega$ near a peak \cite{Removing-Fermi}, it seems from \disp{Eq-1,def-Phi} that we need  more. The Hilbert transform term $\Phi_{\k}(\omega)  $,  
  requires a knowledge of    $A(\k,\nu)$ at {\em all} $\nu$ in order to  determine $\rho_\Sigma$ rigorously. Any error in the estimated $\rho_\Sigma(\k,\omega)$ therefore arises only  from errors in evaluating $\Phi_{\k}(\omega)$ due to  a limited (partial) knowledge of $A(\k,\omega)$.

\begin{figure}[htpb]
\centering
\includegraphics[width=.63\textwidth]{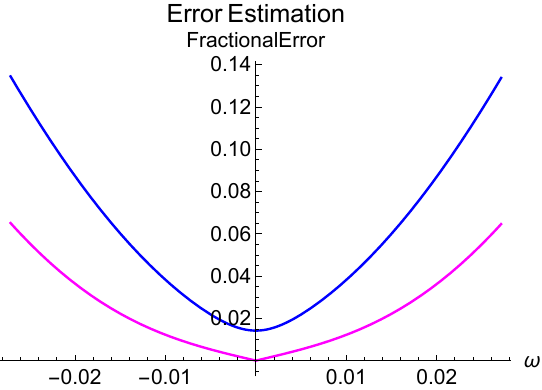}\hfill
\caption{\footnotesize The fractional error \disp{Frac-error-exact} (lower magenta curve) and its upper bound \disp{Frac-error} (upper blue curve) for the Asymmetric Fermi-liquid model \disp{Model-I} with  parameters specified below in \disp{Paras-AFL }. Here $\omega$ is in units of  eV as per the model.  We note that the bound would also vanish at $\omega=0$ if we use a  vanishing  elastic scattering energy $\eta$ in \disp{Paras-AFL}.}
\label{Figure1}
\end{figure}

On closer inspection, we find that the situation is sensitively  dependent on the  regions of $\k,\omega$ probed. We summarize  our observation about achieving highest accuracy in estimating $\rho_\Sigma$ as:
\begin{theorem*}1[{on highest accuracy inversion}]
  As the  energy is lowered to zero,  errors  in  the Hilbert transform term $\Phi_{\k}(\omega)$ are of diminishing  consequence to $\rho_\Sigma(\k,\omega)$.  \label{HAT}
 \end{theorem*}
To understand the origin of this  theorem, consider the following.  If we allow for an error  $\delta \Phi_{\k}(\omega)$ at a fixed $\k$ in the estimation of $\Phi_{\k}(\omega)$, the resulting fractional error in   $\rho_\Sigma(\k,\omega)$   to the first order   is given by
\beq
\frac{1}{\rho_\Sigma(\k,\omega)} \Big\vert \frac{\delta  \rho_\Sigma(\k,\omega)}{\delta \Phi_{\k}(\omega)}\Big\vert&=& 2 \sqrt{\frac{\rho_\Sigma(\k,\omega)}{A(\k,\omega)}}\sqrt{1- \pi^2 A(\k,\omega) \rho_\Sigma(\k,\omega)}\label{Frac-error-exact} \\
&\leq &2 \sqrt{\frac{\rho_\Sigma(\k,\omega)}{A(\k,\omega)}}.
\label{Frac-error}
\eeq
In \figdisp{Figure1} we illustrate the fractional error and its upper bound  given in \disp{Frac-error-exact,Frac-error} for the Asymmetric- Fermi-liquid  model \disp{Model-I} defined below.
For $\k$$\sim$$\k_{F}$ and $\omega$$\sim$$0$, the spectral function $A(\k,\omega)$ has a peak,  and  the self-energy term $\rho_\Sigma(\k,\omega)$ is expected to vanish at $T=0$  in disorder-free Fermi systems. The expression \disp{Frac-error-exact} can also be written as $ 2 |\Re\, G(\k,\omega)| \frac{\rho_\Sigma(\k,\omega)}{A(\k,\omega)}$, which provides further understanding of the vanishing at $\omega=0$, in terms of the expected vanishing of $\Re\, G(\k_F,\omega\sim0)$.
 The cumulative effect is that  the fractional error and its upper bound  \disp{Frac-error} are least in the low energy regime. This regime is also the most interesting one from a physical standpoint, since it defines the asymptotic low energy physics of the system, where the behavior of the self-energy is an important characterization  of the physics of the system.   Therefore  the above theorem provides a strong motivation to explore the approximate evaluation of \disp{Eq-1}, as a way to probe fundamental  aspects  of interacting Fermi systems.

Encouraged by the above discussion, we propose that this method for extracting the electron self-energy deserves some experimental effort.
We can state our proposal in qualitative terms as follows: at a fixed $\k$ (chosen say as   $\k_F$),  the low energy behavior of $\rho_\Sigma(\k_F,\omega)$, can be found from a knowledge of $A(\k_F,\omega)$ in a range of energies sufficiently close to  its peak from \disp{Eq-1} supplemented with a suitable frequency windowing  of $A(\k_F,\omega)$. By frequency windowing we mean replacing 
$A(\k_F,\omega)$ in \disp{Eq-1} as  
\beq
A(\k_F,\omega)\to A'(\k_F,\omega) \equiv  {\cal T}(\omega) A(\k_F,\omega),
\label{cutoff} \eeq 
 and ${\cal T}(\omega)$ is a smooth symmetric function of $\omega$, falling from 1 to zero smoothly beyond a suitable cutoff energy $\AW$. Examples of useful forms of ${\cal T}(\omega)$ are provided below.  
This procedure can also be carried out  for arbitrary $\k$ away from $\k_F$  with
 $A'(\k,\omega)={\cal T}(\omega-\omega_{\k}) A(\k,\omega)$
where $\omega_k$ is the location of the peak in $A(\k,\omega)$.

Clearly the proposal is not  rigorous. The  possibility of higher order terms in $\delta \Phi_{\k}$  becoming dominant must be kept in mind while drawing conclusions from the above linear analysis based Theorem. We provide numerical examples below that address this aspect of the problem. The examples displayed below  suggest that in certain cases, the procedure leads to a reasonable reconstruction in the low energy regime where it is  increasingly accurate- in accordance with the Theorem. In a variety of physically interesting  examples, we start with a  self-energy and construct the spectral function from it. We then use the spectral functions cutoff at some energy scale, following the lines suggested in  the proposal, from which  we reconstruct the self-energy. Comparing the reconstructed and original  selfenergies gives us useful insights. In the examples that  are provided,  it seems  that the presence of a sharp peak in $A(\k,\omega)$ is helpful, as the Theorem-1 suggests. The utilization of the values of $A(\k,\omega)$  in a range of energies around the peak discussed below, yields an excellent picture of the low energy behavior of the self-energy.

\subsection{ The cutoff functions:}
 We suggested above the use of a windowed version of $A(\k,\omega)$ as in \disp{cutoff}. In choosing to  cutoff the frequency at a specific $\AW$, typically $\AW \sim \nu \AAW$ i.e. a few times the width of the spectral peak $W_0$,  we must specify the cutoff function ${\cal T}(\omega)$. In  taking the Hilbert transform, it seems useful to consider alternate forms of the cutoff. 
 
 We  tested  a sharp cutoff function
 \beq
 {\cal T}_{Hard}(\omega) =\Theta(\AW- |\omega|) \label{Hardcutoff}
 \eeq
 where $\AW$ is the cutoff energy.
 We also  used a cutoff function, inspired by  the Tukey window in Fourier transforms,  that seems more promising.  This piecewise function determined by the cutoff scale $\AW$ and {\em two} positive numbers $\nu_{-} < \nu_{+}$, in terms of which we define
\beq
{\cal T}(\omega)\Big\vert_{\{ \nu_{-} \AAW,\nu_{+}\AAW\}}&=& 1 \; \mbox{ for  } |\omega|< \nu_{-} \AAW \nn \\
&=&\half \left(1+ \sin \frac{\pi}{2} \left\{ \frac{1+\nu_{+} - 2 |\omega|/\AAW}{\nu_{+}-1} \right\} \right) \; \nn \\
&&\mbox{ for  }  \nu_{-} \AAW\, < |\omega| < \nu_{+} \AAW \nn\\
&=&0  \mbox{ for  } \nu_{+} \AAW \leq |\omega|. \label{Tukey}
\eeq
This function is displayed in \figdisp{Figure2}.
The scale $\AAW$  is taken in most figures as the width (FWHM) of the spectral function,  typical values of the $\nu$ numbers are $\nu_{-}  \sim3 $ and $\nu_{+} \sim 6 $.  Results for both windows are compared below in \figdisp{Figure4}. While they agree at the lowest energies, in accord with the Theorem, the comparison suggests that the window in \disp{Tukey} is somewhat better as we go away from $\omega=0$.  Without making any claim to  its   being optimal,  we only use the  window in \disp{Tukey}  for  further results. 


\section{Examples  of self-energy inversion in three model systems \label{Examples}}
In order to explore this problem of partial range reconstruction, we study three self energy models next. We also set $\k=\k_F$ in this part of the work, this is the simplest case where $\Delta k$ in \disp{def-Deltak} and \disp{spectral-fromself-energy-1} can be set at zero. Calculations for  $\k\neq\k_F$  can be done in exactly the same way, by shifting the peak in the cutoff function \disp{Tukey} from 0 to $\delta \k$, the location of the  spectral function peak. We comment on this in \figdisp{Figure2}.
Each  model is  given in terms of
an explicit positive definite self energy  $\rho_\Sigma(\k_F,\omega)$ dependent on a few parameters, and having a  finite integral over $\omega$. 

The   three illustrative  models considered  are expressed (see \disp{AFL-Paras-list}) in terms of dimensionless (scaled) frequency $\bw\equiv \frac{\omega}{\Omega_0}$,  a dimensionless temperature $\bar{\tau}= \frac{ \pi k_B T}{\Omega_0}$, a dimensionless interaction strength parameter $\bar{\epsilon}_0=\frac{\epsilon_0}{\Omega_0}$.  In the case of the first model we also use $\bA=\frac{A}{\Omega_0}$, it is  a dimensionless asymmetry parameter. We also use the dimensionless version of $\rho_\Sigma$
\beq
\bar{\rho}_\Sigma(\k_F,\bw)= \frac{1}{\Omega_0} {\rho}_\Sigma(\k_F,\omega)
\eeq
where $\Omega_0$ is the ``large energy'' scale. 
The relationship between  $\Omega_0$ and the spectral peak width $W_0$ is determined by  uninteresting details of the model used. For example changing the confining well from the Gaussian to another form in \disp{Model-I,Model-II-a,Model-II-b,Model-III} would change that relation. Therefore we choose  $\Omega_0=1$eV and adjust other parameters so that the experimentally observable width (FWHM)  $W_0$ of the peak at $\omega\sim0$ is about 10 meV.
 These peak widths seem to be typical values for the scales  for many high T$_c$ materials \cite{Sobota,Valla}.
The three models, chosen for their proximity to  interesting physical cases as well as for analytical tractability in a few cases, 
 are defined by three choices of $\bar{\rho}_\Sigma$
\beq
\mbox{(Asymmetric FL) } \bar{\rho}_\Sigma(\k_F,\bw) &=& {\epsilon_0} \left(\btau^2+\bw^2\right) \left(1-\alpha \bw \right)
   e^{-\bw^2} \label{Model-I} \\
   \mbox{(Marginal-FL-a) } \bar{\rho}_\Sigma(\k_F,\bw) &=& {\epsilon_0} \mbox{Max}[\btau, |\bw|]\; 
   e^{-\bw^2}   \label{Model-II-a} \\
   \mbox{(Marginal-FL-b) } \bar{\rho}_\Sigma(\k_F,\bw) &=& {\epsilon_0} \left( \btau+ |\bw| \right)\; 
   e^{-\bw^2}   \label{Model-II-b} \\
\mbox{(Non-FL) } \bar{\rho}_\Sigma(\k_F,\bw) &=& {\epsilon_0} \left(\btau^\frac{3}{2}+|\bw|^\frac{3}{2}\right) 
   e^{-\bw^2}.   \label{Model-III} 
   \eeq
   These models are a non-exhaustive subset of the models discussed in literature, and chosen to provide a fairly broad diversity of behaviour.
 The low frequency behaviour of these models are of especial interest, where the Gaussian term $e^{-\bw^2}$ is essentially unity. This  term is a ``confining well'',  chosen to provide a fall off at high $\bw$ needed for the integrability of the spectral density. Other choices of the confining well are  possible but unlikely to make a difference at  low $\bw\ll1$, which is the region of our main concern. 
 
 In \disp{Model-I} the chosen  Asymmetric FL function (A-FL in the following) describes an asymmetric Fermi liquid, where the first term $(\bar{\tau}^2+\bw^2)$ represents a Fermi liquid (FL), while the second  term $(1-\alpha \bw)$ generates a cubic asymmetric term in the self energy. This  model is a crude representation of  the solution of the \tJ model in 2-dimensions using  the extremely correlated Fermi liquid theory (ECFL)\cite{ECFL-2D}  at the lowest temperatures, if the parameters are chosen appropriately.
The resistivity of this model is quadratic in T at the lowest temperature, and crosses over to a T-linear    behavior at a very low crossover temperature (\cite{ECFL-2D}(d)). This subtle crossover  behavior requires the addition of terms of higher order in $\omega$, and  is  buried in the $T$ dependence of the coefficients. These details are not necessary for  the basic analysis here, in this work we will  provide a framework from which  only  the lowest order quadratic behavior- on display in \disp{Model-I}- might be tested in future experiments. The spectral function can be calculated fully in terms of the Dawson function $D_F$ \disp{Dawson-Def}, these and other necessary details are collected in Appendix(\ref{app-2}).

In \disp{Model-II-a,Model-II-b} we consider two variants of the popular marginal Fermi liquid (M-FL in the following)  model self energy\cite{MFL}. As in other models considered, the  added exponential term is unity  for $|\bw|\ll1$. The M-FL phenomenology builds in a T linear resistivity in a natural fashion.  The model M-FL-b \disp{Model-II-b}  is in the same spirit as model M-FL-a  \disp{Model-II-a}, and leads to slightly different results at the lowest energy.

In \disp{Model-III} the chosen non FL function (N-FL in the following) describes a non-Fermi liquid system with a power law $|\omega|^{\frac{3}{2}}$, where the resistivity is expected to behave as $T^{\frac{3}{2}}$. While there appears to be no compelling argument  
for this specific choice of the power law ${\frac{3}{2}}$ used in our choice, we use it as an archetype of a strongly non Fermi liquid of the type suggested in \cite{SYK}.

   We adjust the parameters for these models so that the spectral function has about the same width $W_0$  of $\sim 10$ meV  in all cases.

 In summary, in the following we follow these steps: 
 
 \begin{itemize}  
\item  For each model we use \disp{chi-def,spectral-fromself-energy-1} to construct  $\chi$, and $A(\k,\omega)$. 
\item  We then multiply  $A$ with a  window function ${\cal T}(\omega)$ in \disp{Tukey}, which vanishes smoothly beyond a frequency scale. This scale is chosen in most cases as $\AW$,  a suitable multiple of $W_0$, the width (FWHM) of the spectral function. This process  represents the selection of a small energy window and yields $A'$ as in \disp{cutoff}.
\item From  $A'$ replacing $A$  and \disp{KK,self-energy-fromspectral-11}, we construct $\rho_\Sigma(\k,\omega,\AW)$, which now depends on our choice of $\AW$. We then compare with the parent value $\rho_\Sigma(\k_F,\omega)$, for $|\omega|\lessim \AW$. 
\end{itemize}
As a  check on the numerics and the formalism, we note that   the two self energies must agree when $\AW\!\to\! \infty$.  We present an example in \figdisp{Figure2} to demonstrate this agreement.

\subsection{ The Asymmetric Fermi liquid model}
 The parameters defined in \disp{AFL-Paras-list} are  chosen for the following figures are 
\beq
\epsilon_0=1.8, \; \alpha=0.1; \; \eta=0.02; \; \tau=0.02; \; \alpha_z= 1.74  \label{Paras-AFL}
\eeq
where $\alpha_z$ \disp{Gutzwiller} corresponds to a filling $n=0.85$,
leading to a fairly low value of the quasiparticle weight $Z=0.203$. The energies are given in units of $\Omega_0$ chosen to be  1eV  for  High Tc systems.
The value of $\eta$ used here  corresponds to typical Laser ARPES experiments \cite{Gweon}, while $\tau$,  the physical temperature $T=\tau/\pi\sim 74$K. The  width (FWHM) of the spectral function at these parameters is $W_0$ is $\sim$ 9 meV.

\begin{figure}[htpb]
\centering
\includegraphics[width=.36\textwidth]{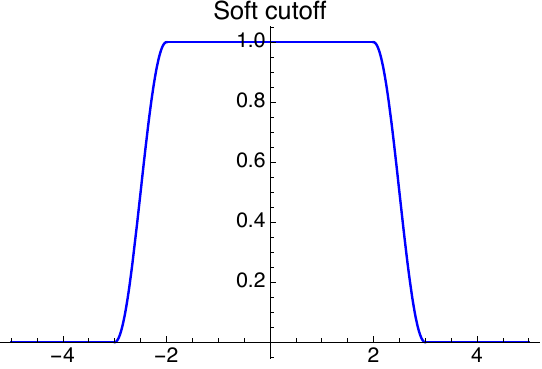}
\includegraphics[width=.54\textwidth]{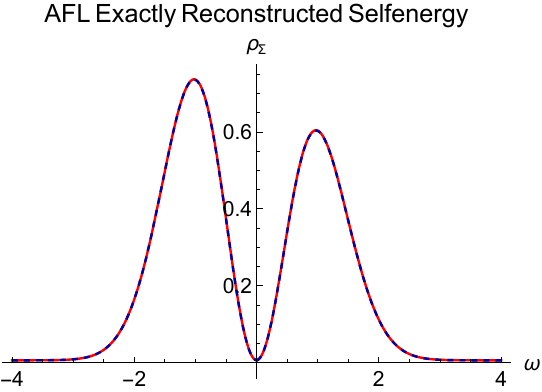}
\caption{\footnotesize {These figures are   for the Asymmetric Fermi-liquid model \disp{Model-I} with  parameters given  in \disp{Paras-AFL }. \bf Left} The soft cutoff  function ${\cal T}(\omega)$ in \disp{Tukey} is illustrated  with parameters $\{ 2,3\}$ and $\AW=1$. {\bf Right} The initial $\rho_\Sigma(\k_F,\omega)+\frac{\eta}{\pi}$ from \disp{rhototal,expandself-energy} (dotted blue) is exactly reproduced numerically (red), using the full energy window Hilbert transform to evaluate
$\dashint d\nu \frac{A(\k,\nu)}{\omega-\nu}$ in \disp{Eq-1}. It also agrees with the analytical calculation in \disp{Real-G}, showing the consistency of the starting point \disp{Eq-1}.  A similar calculation performed away from $\k_F$ (as mentioned  in the beginning of Section(\ref{Examples})) works equally well. }
\label{Figure2}
\end{figure}

\begin{figure}[htpb]
\centering
\includegraphics[width=.49\textwidth]{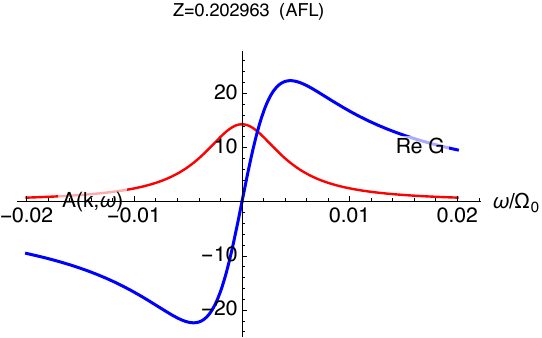}
\includegraphics[width=.49\textwidth]{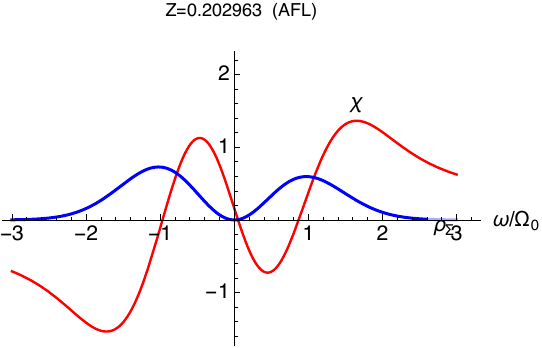}
\caption{\footnotesize These figures are   for the Asymmetric Fermi-liquid model \disp{Model-I} with  parameters given  in \disp{Paras-AFL }. ({\bf Left}) The spectral function $A(\k_F,\omega)$ and the $Re \, G(\k_F,\omega)$  and ({\bf Right}) the  imaginary self energy $\rho_\Sigma(\k_f,\omega)$  and $\chi(\k_F,\omega)$ for the A-FL model \disp{Model-I}, calculated from \disp{chi,expandself-energy,spectral-function-scaled}, using parameters given in \disp{Paras-AFL}. Here the spectral peak width $W_0$ is $\sim$9 meV.}
\label{Figure3}
\end{figure}
We present figures showing the spectral function and self-energy for the A-FL model in \figdisp{Figure3}. We further
 demonstrate the validity of the basic \disp{Eq-1} for the A-FL  model. In   \figdisp{Figure2} we show  the exact reconstruction of the self-energy from the spectral function using the Hilbert transform over all frequencies.

\begin{figure}[htpb]
\centering
\includegraphics[width=.49\textwidth]{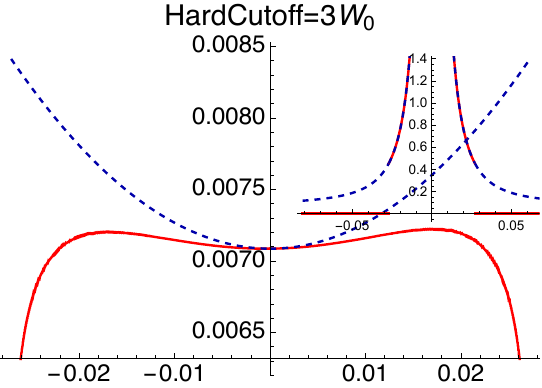}
\includegraphics[width=.49\textwidth]{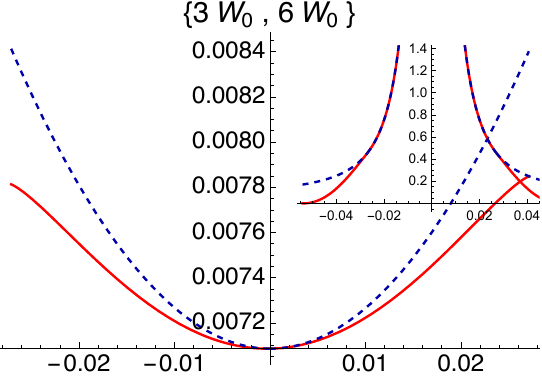}
\caption{\footnotesize  These figures are   for the Asymmetric Fermi-liquid model \disp{Model-I} with  parameters given  in \disp{Paras-AFL }. The reconstructed self-energy $\rho_\Sigma(\k_F,\omega)+\frac{\eta}{\pi}$ (red curve) using on ({\bf Left}) a hard cutoff \disp{Hardcutoff} and on ({\bf Right}) using the soft cutoff \disp{Tukey}  compared with the exact value in the dashed blue curves.
The two  parameters relating to the soft cutoff \disp{Tukey} on ({\bf Right})  are indicated in curly brackets.
  The two  insets show the exact spectral function in dashed blue and the cutoff included spectral function $A'$ of \disp{cutoff}  in the red curves. Here $W_0$ is  $\sim$9 meV.  On the ({\bf Left}), while the hard cutoff does give a shallow minimum near $\omega$$\sim$$0$, it is seen to turn around and become convex rapidly. The soft cutoff remains concave up to $3 W_0$, and has a maximum (fractional) error of $\sim$6\% at the maximum displayed energy  $\omega=3 W_0$.  }
\label{Figure4}
\end{figure}
\begin{figure}[htpb]
\centering
\includegraphics[width=.31\textwidth]{AFLfig36.pdf}
\includegraphics[width=.31\textwidth]{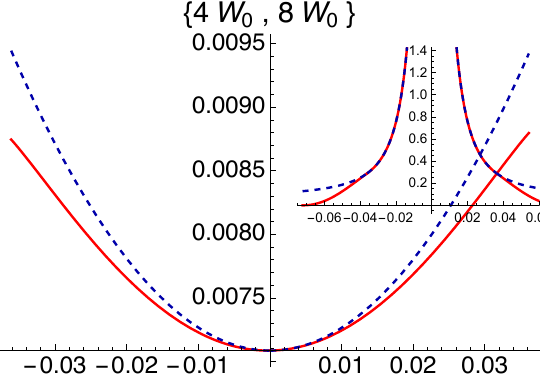}
\includegraphics[width=.31\textwidth]{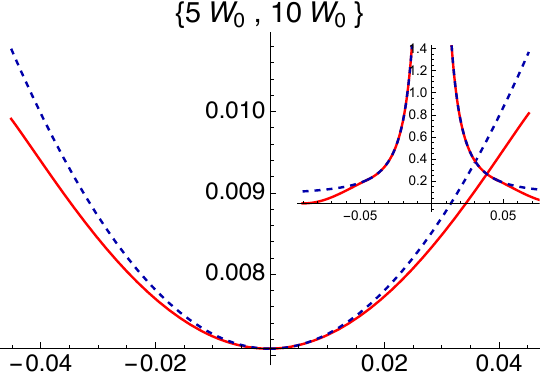}
\caption{\footnotesize These figures are   for the Asymmetric Fermi-liquid model \disp{Model-I} with  parameters given  in \disp{Paras-AFL }. The reconstructed self-energy $\rho_\Sigma(\k_F,\omega)+\frac{\eta}{\pi}$ (red curves)  using the soft cutoff \disp{Tukey}
 compared with the exact value in the dashed blue curves.
 The two  parameters relating to the soft cutoff \disp{Tukey} are indicated in curly brackets.
  The three  insets show the exact spectral function in dashed blue and the cutoff included spectral function $A'$ of \disp{cutoff}  in the red curves. Here upper cutoff is twice the lower one. Increasing the magnitude of the cutoff is seen to reduce the error.  Here $W_0$ is  $\sim$9 meV.
}
\label{Figure5}
\end{figure}
\begin{figure}[htpb]
\centering
\includegraphics[width=.31\textwidth]{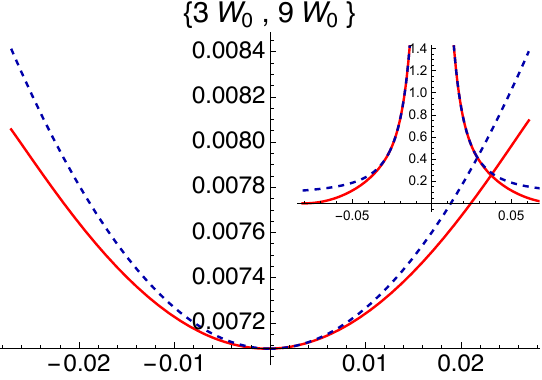}
\includegraphics[width=.31\textwidth]{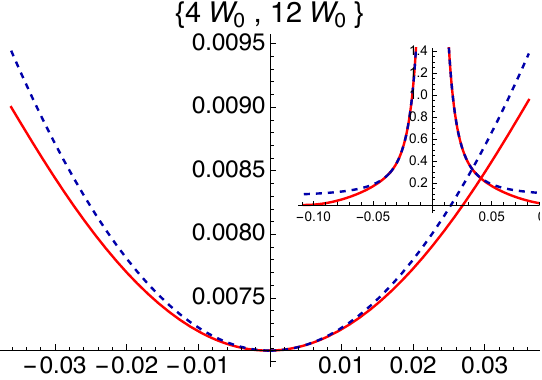}
\includegraphics[width=.31\textwidth]{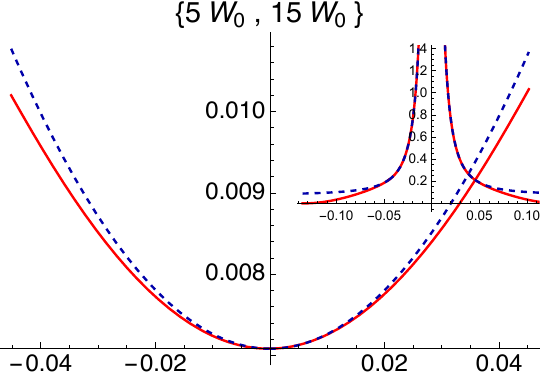}
\caption{\footnotesize These figures are   for the Asymmetric Fermi-liquid model \disp{Model-I} with  parameters given  in \disp{Paras-AFL }. The reconstructed self-energy $\rho_\Sigma(\k_F,\omega)+\frac{\eta}{\pi}$ (red curves)  using the soft cutoff \disp{Tukey}
 compared with the exact value in the dashed blue curves.
 The two  parameters relating to the soft cutoff \disp{Tukey} are indicated in curly brackets.
  The three  insets show the exact spectral function in dashed blue and the cutoff included spectral function $A'$ of \disp{cutoff}  in the red curves. Here upper cutoff is 3 times the lower one, and shows a fair improvement over the results of \figdisp{Figure5} where the upper cutoff is  only twice the lower one.  Here $W_0$ is  $\sim$9 meV.}
\label{Figure6}
\end{figure}

We present a  comparison  in \figdisp{Figure4}  of the reconstruction schemes using  two window functions: (a) A hard cutoff ${\cal T}(\omega)= \Theta( 3 W_0-|\omega|)$ at a   energy cutoff equaling thrice the FWHM of the spectral peak $W_0$ and (b) the soft window given in \disp{Tukey} with parameters $\{3 W_0, 6 W_0\}$. In the caption we we comment further on the relative merits of the two schemes. In the two figures \figdisp{Figure5,Figure6}, we display the reconstructed self-energy compared to the exact self-energy for different sets of the cutoff window parameters, and comment on their relative merits.

\subsection{ The Marginal Fermi liquid  model}
\begin{figure}[htpb]
\centering
\includegraphics[width=.49\textwidth]{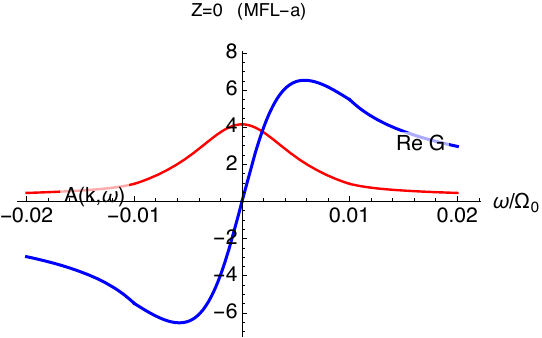}
\includegraphics[width=.49\textwidth]{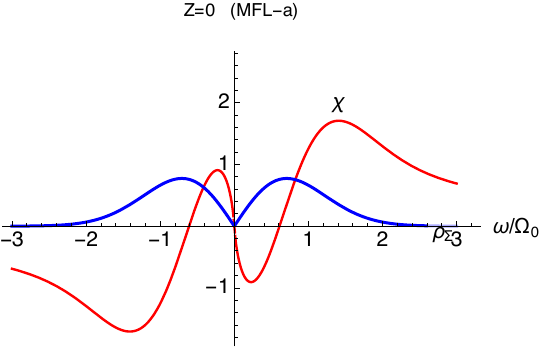}
\caption{\footnotesize These figures are   for the Marginal Fermi Liquid-a  model \disp{Model-II-a} with  parameters given  in \disp{Paras-MFLa }.  ({\bf Left}) The spectral function $A(\k_F,\omega)$ and the $Re \, G(\k_F,\omega)$  and ({\bf Right}) the  imaginary self energy $\rho_\Sigma(\k_f,\omega)$  and $\chi(\k_F,\omega)$ for the M-FL-a model \disp{Model-II-a}, calculated from \disp{chi,expandself-energy,spectral-function-scaled}. Here the spectral peak width $W_0$ is $\sim$12 meV.}
\label{Figure7}
\end{figure}

\begin{figure}[htpb]
\centering
\includegraphics[width=.31\textwidth]{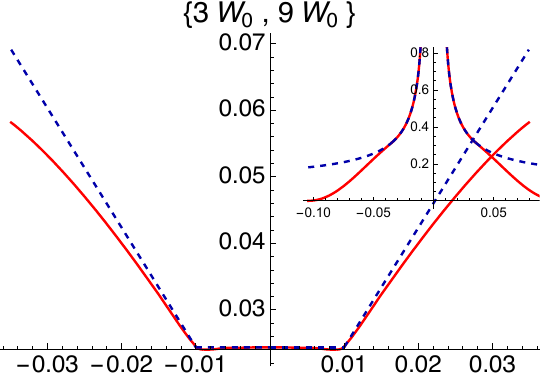}
\includegraphics[width=.31\textwidth]{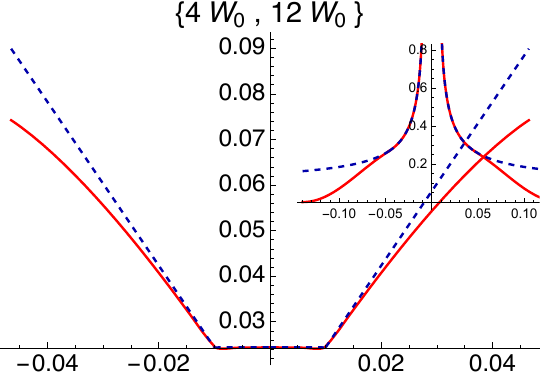}
\includegraphics[width=.31\textwidth]{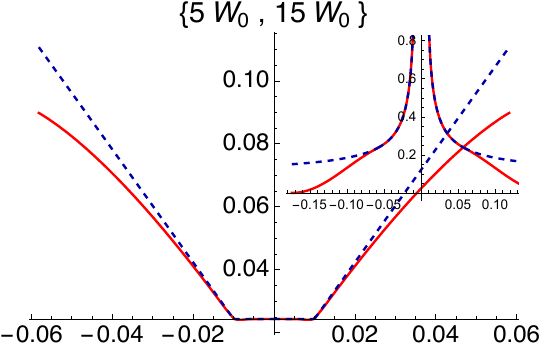}
\caption{\footnotesize  These figures are   for the Marginal Fermi Liquid-a  model \disp{Model-II-a} with  parameters given  in \disp{Paras-MFLa }. The reconstructed self-energy $\rho_\Sigma(\k_F,\omega)+\frac{\eta}{\pi}$ (red curves)  using the soft cutoff \disp{Tukey}
 compared with the exact value in the dashed blue curves.
 The two  parameters relating to the soft cutoff \disp{Tukey} are indicated in curly brackets.
  The three  insets show the exact spectral function in dashed blue and the cutoff included spectral function $A'$ of \disp{cutoff}  in the red curves. Here upper cutoff is 3 times the lower one.  Here $W_0$ is $\sim$12 meV.}
\label{Figure8}
\end{figure}

For the M-FL-a model, the parameters used are 
\beq
\epsilon_0=1.8, \eta=0.02,\tau=0.01; \alpha_z=1 \label{Paras-MFLa}
\eeq
leading to a width $W_0=12$ meV, which is only slightly bigger than $W_0=9$ meV for the A-FL model. The spectral function and the real part of the Greens function  are shown in \figdisp{Figure7},  and
 the reconstructed self-energy with a few typical window parameters also in \figdisp{Figure7}. The reconstructed self-energy is displayed in \figdisp{Figure8}.

\begin{figure}[htpb]
\centering
\includegraphics[width=.49\textwidth]{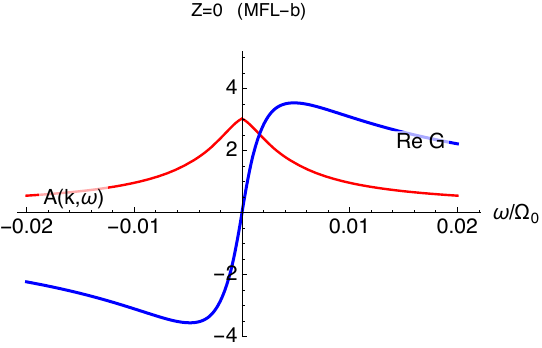}
\includegraphics[width=.49\textwidth]{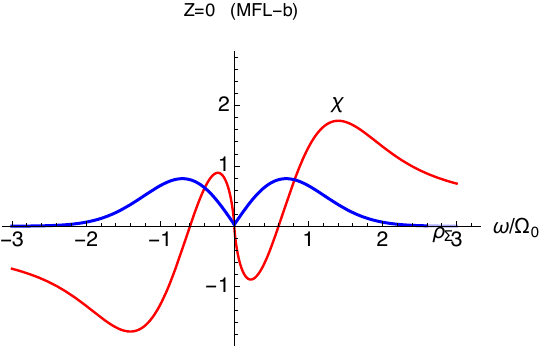}
\caption{\footnotesize These figures are   for the Marginal Fermi Liquid-b  model \disp{Model-II-b} with  parameters given  in \disp{Paras-MFLb }.  ({\bf Left}) The spectral function $A(\k_F,\omega)$ and the $Re \, G(\k_F,\omega)$  and ({\bf Right}) the  imaginary self energy $\rho_\Sigma(\k_f,\omega)$  and $\chi(\k_F,\omega)$,  calculated from \disp{chi,expandself-energy,spectral-function-scaled}, using parameters given in \disp{Paras-MFLb}. Here the spectral peak width $W_0$ is $\sim$11 meV.}
\label{Figure9}
\end{figure}

\begin{figure}[htpb]
\centering
\includegraphics[width=.31\textwidth]{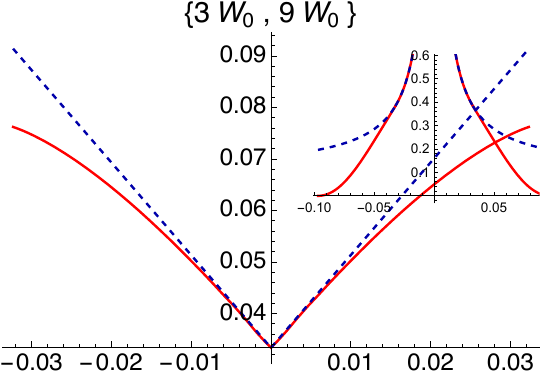}
\includegraphics[width=.31\textwidth]{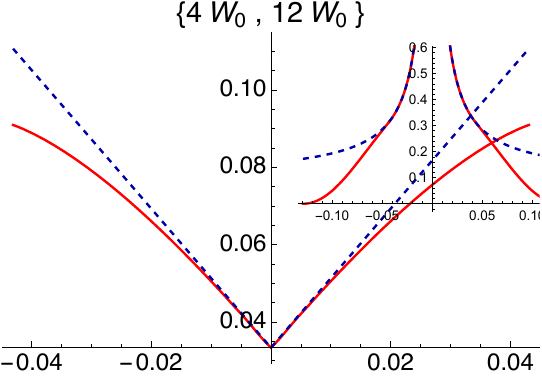}
\includegraphics[width=.31\textwidth]{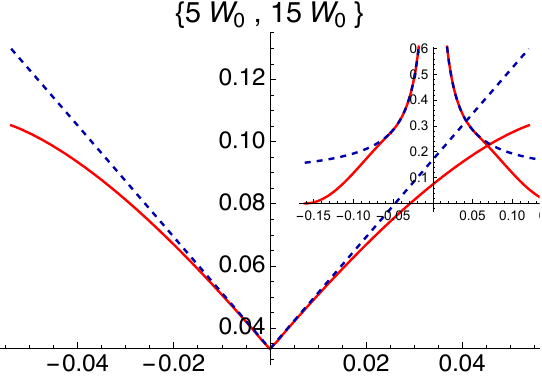}
\caption{\footnotesize These figures are   for the Marginal Fermi Liquid-b  model \disp{Model-II-b} with  parameters given  in \disp{Paras-MFLb }.  The reconstructed self-energy $\rho_\Sigma(\k_F,\omega)+\frac{\eta}{\pi}$ (red curves)  using the soft cutoff \disp{Tukey}
 compared with the exact value in the dashed blue curves.
 The two  parameters relating to the soft cutoff \disp{Tukey} are indicated in curly brackets.
  The three  insets show the exact spectral function in dashed blue and the cutoff included spectral function $A'$ of \disp{cutoff}  in the red curves. Here upper cutoff is 3 times the lower one.  Here $W_0$ is $\sim$11 meV. Note the difference in the shape at low $\omega$ from that in \figdisp{Figure8} This version has a linear behaviour down to the lowest energy.}
\label{Figure10}
\end{figure}

The functional form of $\rho_\Sigma$ in the M-FL-a model \disp{Model-II-a} has a flat portion in its minima, which is reflected in the lowest energy behaviour as seen  in \figdisp{Figure8}. The M-FL-b model \disp{Model-II-b} on the other hand avoids this flat feature.
For the M-FL-b model, the parameters used are 
\beq
\epsilon_0=1.8, \eta=0.02,\tau=0.015; \alpha_z=1 \label{Paras-MFLb}
\eeq
leading to a width $W_0=11$ meV.
We now show the  spectral function and the real part of $G$ in \figdisp{Figure9} and the reconstructed self-energy in  \figdisp{Figure10} for the M-FL-b model given by \disp{Model-II-b}, with a few typical window parameters.

\subsection{A Non Fermi liquid model}
\begin{figure}[htpb]
\centering
\includegraphics[width=.49\textwidth]{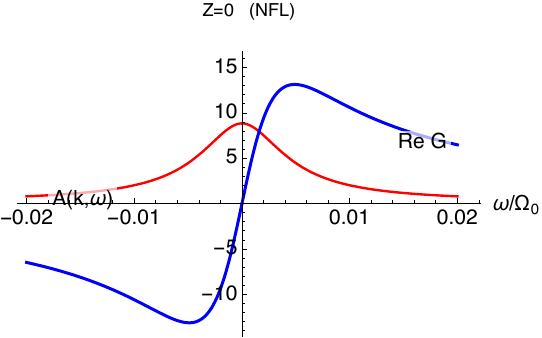}
\includegraphics[width=.49\textwidth]{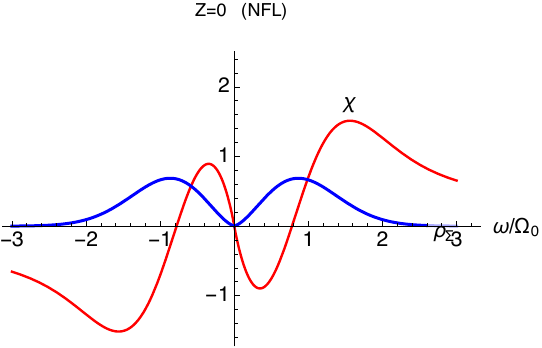}
\caption{\footnotesize   These figures are   for the Non Fermi Liquid  model \disp{Model-III} with  parameters given  in \disp{Paras-NFL }.   ({\bf Left}) The spectral function $A(\k_F,\omega)$ and the $Re \, G(\k_F,\omega)$  and ({\bf Right}) the  imaginary self energy $\rho_\Sigma(\k_f,\omega)$  and $\chi(\k_F,\omega)$, calculated from \disp{chi,expandself-energy,spectral-function-scaled}, using parameters given in \disp{Paras-NFL}. Here the spectral peak width $W_0$ is $\sim$10 meV.}
\label{Figure11}
\end{figure}

\begin{figure}[htpb]
\centering
\includegraphics[width=.31\textwidth]{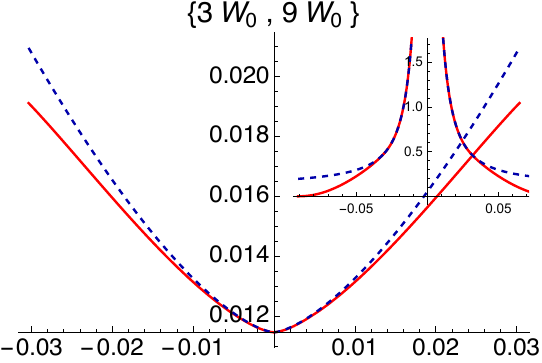}
\includegraphics[width=.31\textwidth]{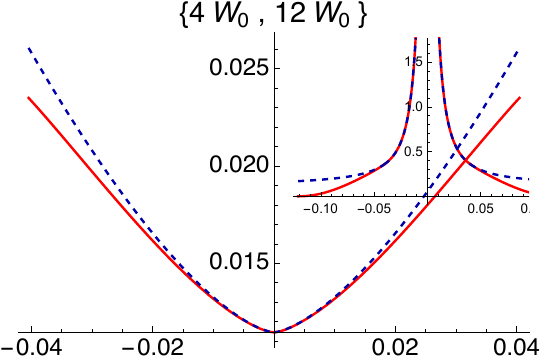}
\includegraphics[width=.31\textwidth]{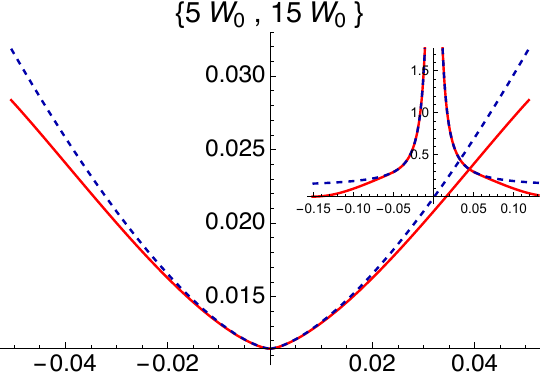}
\caption{\footnotesize These figures are   for the Non Fermi Liquid  model \disp{Model-III} with  parameters given  in \disp{Paras-NFL }.   The reconstructed self-energy $\rho_\Sigma(\k_F,\omega)+\frac{\eta}{\pi}$ (red curves)  using the soft cutoff \disp{Tukey}
 compared with the exact value in the dashed blue curves.
 The two  parameters relating to the soft cutoff \disp{Tukey} are indicated in curly brackets.
  The three  insets show the exact spectral function in dashed blue and the cutoff included spectral function $A'$ of \disp{cutoff}  in the red curves. Here upper cutoff is 3 times the lower one. Here $W_0$ is $\sim$10 meV. }
\label{Figure12}
\end{figure}

For the N-FL model \disp{Model-III}, the parameters used are 
\beq
\epsilon_0=1.8, \eta=0.02,\tau=0.02; \alpha_z=1 \label{Paras-NFL}
\eeq
leading to a width $W_0=10$ meV. 
We next display the  spectral function and the real part of $G$ in \figdisp{Figure11} and the reconstructed self-energy \figdisp{Figure12} for the N-FL model given by \disp{Model-III}, with a few typical window parameters.

We now show the reconstructed self-energy with a few typical window parameters




\section{Comments and Conclusions \label{Conclusions}}

 The proposal for reconstructing the self-energy from the spectral function made in this work in Section(\ref{proposal}), was illustrated above in Section(\ref{Examples}) in a set  of figures (\figdisp{Figure3}-\figdisp{Figure12}) using three typical models with different predictions.
 These figures show that an experimental implementation of the proposal could lead to interesting insights about the nature of quantum matter.

Different theoretical approaches to the strong correlation problem, originally inspired by the High $T_c$ cuprates but branching out to a much broader portfolio of materials in recent years, lead to a variety of different  self energies- some of them are discussed in this work.
Fundamental questions about the nature of these  quantum materials remain open in most cases- it is unclear  whether they constitute some variety of Fermi liquids, or  some variety of non Fermi liquids. While the resistivity is often used to discriminate between these states of matter,
it is a much more complex probe to interpret robustly. 
On the other hand   a much more direct avenue  to answer the above basic question, is to study the (imaginary part of the) self-energy  at low energies.  At present it seems that no decisive tests using experimental ARPES data has been carried out in that direction. This is the motivation for the present paper, where we  present a method to extract the  low $\omega$ self energy from ARPES derived spectral function.  It is based on the helpful Theorem-1, which assures us that the inversion method used  has least error at low energies. 

The proposal presented in Section(\ref{proposal}) is to use \disp{Eq-1} and a suitable windowing of the spectral function, analogous to that in \disp{Tukey} to infer the imaginary self energy from the ARPES  derived spectral function. We provide several examples in Section({\ref{Examples}) that show that the different  model systems yield distinguishably different low energy selfenergies.

We conclude with a few comments
\begin{itemize}

\item[(1)] The suggested inversion process can be used to estimate the elastic scattering parameter $\eta$, from the T independent part of 
the derived self-energy as $\omega\to0$. \change{ For this purpose one can use the inversion data at a few (typically two or three) distinct temperatures together  with \disp{rhoscaled,rhototal} to deduce $\eta$.}

\item[(2)] The observed peaks in the spectral function at $\k_F$ are expected to have a T dependent shift  given in \disp{def-Deltaw,delta-omega-unscaled,delta-omega-scaled}. The shift in \disp{delta-omega-scaled}
contains the chemical potential part that can be estimated from the thermopower.  The remainder has two terms in it, but is expected to be dominated by the term containing the asymmetry parameter $\alpha$ (see \disp{Model-I}). This can be a useful way to estimate $\alpha$ from the low energy experiments.

\item[(3)] The role of noise in the spectral function warrants  mention; the  Theorem-1  also applies to the noise. This offers hope that at low energies, the errors due to noise are least. 

\item[(4)] While we have focussed on $\k=\k_F$ here, it might be possible to explore departures from the Fermi surface if high quality spectral functions are obtainable from the intensities over a wide range of energies. \change{  We are suggesting that $\k$, chosen   in the proximity of   $\k_F$, should suffice for estimating the crucial frequency dependence  $\rho_\Sigma(\k,\omega)$ near $\k_F$. }

 \item[(5)]  The method described here requires the self-energy to be diagonal in spin-space. It   does not extend to arbitrary $\vec{k}$ in  a   superconducting state, where the self-energy is a $2\times2$ matrix on account of  the  anomalous (i.e. pairing) part. However an important  exception is the nodal direction in 2-d   d-wave superconductor, where  the off-diagonal matrix elements of the self-energy vanish,  and hence our inversion  method is  applicable. 
\end{itemize}

\appendices
\section{Summary of Basic Definitions \label{app-1}}
In this section we reorganize the familiar properties of the Greens function \cite{AGD,Nozieres,Fetter} to define specific quantities used in our analysis.

We start with the standard  expression for the retarded Greens function  $G^{-1}(\k,\omega_c)={ \omega_c }   + \mu -\varepsilon_k - \Sigma(\k,\omega_c)$, with  $\omega_c=\omega+ i 0^+$ \cite{AGD,Fetter,Nozieres}. Emphasizing the role of the spectral density of the self-energy $\rho_\Sigma(\k,\omega)= - \frac{1}{\pi} \Im \, \Sigma(\k,\omega_c)$,  we decompose
\beq\Sigma(k,\omega_c) &= &(1- \alpha_z)  (\omega+\mu)+ \Sigma_*(\k)- i \pi  \rho_\Sigma(\k, \omega)+ \chi(\k,\omega)\nn \\
\eeq
where $\chi$ can be obtained from   the Hilbert transform of $\rho_\Sigma(\k,\omega)$ as
\beq
\chi(\k,\omega)= \dashint d\nu  \frac{\rho_\Sigma(\k,\nu)}{\omega-\nu}. \label{chi-def}
\eeq 
By definition $\lim_{\omega\to \infty} \chi(\k,\omega)\to 0$ (assuming an integrable $\rho_\Sigma$). Here $\Sigma_*$ is the static part of the self-energy, analogous to the Hartree-Fock term, which remains finite as $\omega\to\infty$. It   cannot be deduced from $\rho_\Sigma(\k,\omega)$, we see below that it can be absorbed in the measurable shifts of the peaks as defined below in \disp{def-Deltaw,def-Deltak}. The constant $\alpha_Z$ is given by
 \beq
 \alpha_z&=&1, \mbox{   Canonical (Hubbard type) Fermions} \label{Canonical}\\ \alpha_z&=& \frac{1}{1-\half n}, \mbox{ Gutzwiller projected (\tJ type) Fermions} \label{Gutzwiller}
 \eeq
 and reflects the basic nature of the Fermions  in various models \cite{origin-of-alphaz}.
   Using this decomposition we write
\beq
G^{-1}(\k,\omega_c)&=&\alpha_z ({ \omega_c } +\mu)  -\varepsilon_{\k} - \Sigma_*(\k)-\chi(\k,\omega)+ i \pi \rho_{\Sigma}(\k,\omega).\nn \\\label{Ginv-def}
\eeq
We rewrite this expression using the basic idea that at $T=0$ on the Fermi surface $G(\k_F,\omega)$ has a pole at $\omega=0$. Since $\rho_\Sigma(\k_F,0)$ vanishes,  
\beq
\alpha_z \times \mu\vert_{T=0}= \varepsilon_{\k_F} + \left( \Sigma_*(\k_F)+\chi(\k_F,0)\right)\vert_{T=0},
\eeq
which can be used in \disp{Ginv-def} to rewrite it as
\beq
G^{-1}(\k,\omega_c)&=&{ \omega_c } \alpha_z + \Delta\omega- \Delta k+\chi(\k,0)-\chi(\k,\omega)+ i \pi \rho_{\Sigma}(\k,\omega).\nn \\\label{Ginv-def-2}
\eeq
where the real functions $\Delta\omega$ the energy shift, and $\Delta k$ the momentum shift are give by
\beq
\Delta\omega&=& \alpha_z (\mu-\mu\wert)+(\Sigma_*(\k_F)\wert-\Sigma_*(\k_F)) +(\chi(\k_F,0)\wert-\chi(k_F,0)) \nn \\
\label{def-Deltaw}\\
\Delta k&=&( \varepsilon_{\k}-\varepsilon_{\k_F} )+( \Sigma_*(\k)-\Sigma_*(\k_F) )+(\chi(\k,0)-\chi(\k_F,0)) \label{def-Deltak}
\eeq
By their definitions, $\Delta \omega$ vanishes at T=0, while $\Delta k$ vanishes at $\k_F$.

At low T $\Delta \omega$  is expected to be small  and may be estimated from the thermopower $S$  using the approximate Kelvin relation for thermopower  $S_{Kelvin}= \frac{-1}{q_e} \left(\frac{\partial \mu(T)}{\partial T} \right)\Big\vert_{N,V}$, where $q_e$ is the electron charge.

The momentum shift  $\Delta k$ (at arbitrary $T$)  is of ${\cal O}(|\k-\k_F|)$ for $\k$ near  $\k_F$.
We now split $G(\omega+i0^+)$ into its  real and imaginary parts as   
\beq
G(\k,\omega_c)&=& \Re\, G(\k,\omega) -i \pi A(\k,\omega)  \label{re-im-G}
\eeq
so that the spectral function $A(\k,\omega)= - \frac{1}{\pi} \Im \, G(\k,\omega)$. Using \disp{Ginv-def-2}  we get 
\beq
A(\k,\omega)= \frac{\rho_{\Sigma}(\k,\omega)}{ \{\alpha_z \omega +\Delta\omega-\Delta k+\chi(\k,0)-\chi(\k,\omega) )\}^2+ \pi^2 \rho^2_{\Sigma}(\k,\omega)}, \label{spectral-fromself-energy-1}
\eeq
and
\beq
\Re \, G(\k,\omega)= \frac{\{\alpha_z \omega +\Delta\omega-\Delta k+\chi(\k,0)-\chi(\k,\omega) )\}}{ \{\alpha_z \omega +\Delta\omega-\Delta k+\chi(\k,0)-\chi(\k,\omega) )\}^2+ \pi^2 \rho^2_{\Sigma}(\k,\omega)}. \label{Real-G}
\eeq

We now recall that the real and imaginary parts of the (causal) retarded Greens function   $G(\omega+i0^+)$ are also  related by the Kramers-Kronig relation 
\beq
\Re\, G(\k,\omega)= \dashint d\nu \frac{A(\k,\nu)}{\omega-\nu}, \label{KK}
\eeq
so that a complete knowledge of $A$ suffices to determine $\Re\, G$.

Now moving in a slightly different direction,   taking the imaginary part of \disp{Ginv-def} we get
\beq
\rho_\Sigma(\k,\omega)= \frac{1}{\pi}\Im \,G^{-1}(\k,\omega).
\eeq
Using \disp{re-im-G} this give $\rho_\Sigma$ in terms of $A(\k,\omega)$ and $\Re G(\k,\omega)$ as
\beq
\rho_\Sigma(\k,\omega)=\frac{ A(\k,\omega)}{\{\Re\, G(\k,\omega)\}^2+  \{ \pi A(\k,\omega) \}^2} \label{self-energy-fromspectral-11}
\eeq
This equation asserts that if we know the spectral function $A(\k,\omega)$ {\em for  all} $\omega$, we can retrieve  $\rho_\Sigma(\k,\omega)$, since   the   $\Re \, G$  can be inferred through the Hilbert transform \disp{KK}.  Relabeling $\Re \, G$ as $\Phi_{\k}(\omega)$   gives  \disp{Eq-1}.

\section{Details of the Asymmetric Fermi Liquid Model  \label{app-2}}
\subsection{Scaled variables and  expressions for Hilbert transforms}
Let us consider a FL type model with a cubic asymmetry:
We denote 
\beq
\bar{\omega}=\frac{\omega}{\Omega_0},  \;\;\bar{\tau}= \frac{ \pi k_B T}{\Omega_0},\;\;\bA=\frac{A}{\Omega_0}, \;\; \bar{\epsilon}_0=\frac{\epsilon_0}{\Omega_0} \label{AFL-Paras-list}
\eeq


\beq
\rho_\Sigma(\k_F,\omega,T)&=&{\epsilon_0} \left(\btau^2+\bw^2\right) \left(1-\alpha \bw \right)
   e^{-\bw^2} \label{modelA}.
\eeq
We scale the self energy with $\Omega_0$ so that
\beq
\bar{\rho}_\Sigma(\k_F,\bw,\bar{\tau})={\bar{\epsilon}_0} \left(\btau^2+\bw^2\right) \left(1-\alpha \bw \right)
   e^{-\bw^2} \label{rho-Sigma} \label{rhoscaled}
\eeq
In this and also in other models  we implicitly add an elastic scattering term $\eta/\pi$ to $\rho_\Sigma(\k,\omega)$,
\beq
\rho_\Sigma(\k,\omega) \Big\vert_{Total}= \rho_\Sigma(\k,\omega)+ \frac{\eta}{\pi}, \label{rhototal}
\eeq
 as shown in \disp{spectral-function-scaled}. This term  arises from impurity scattering \cite{Elihu}, and is found to be useful in distinguishing between  ARPES at different incident photon energies \cite{Gweon}.  Its corresponding real part arising from 
 causality, is dropped since the bandwidth for this term is very large- typically a few eV's.  In practical terms adding $\eta$ is equivalent to increasing the physical temperature $T$ to $\sqrt{T^2+ \frac{\eta}{\pi^3 \epsilon_0}}$, usually this is a small effect. 
 
In summary the model has the following parameters
\begin{itemize}
\item
$\mbox{ strength of self energy: } \epsilon_0$
\item $\mbox{ reduced temperature: }\btau$
\item $\mbox{ cubic asymmetry: }\alpha$
 \item $\mbox{ large energy scale: }\Omega_0
 $
  \item $\mbox{ elastic scattering scale: }\bar{\eta} $
\end{itemize}
We rewrite \disp{rho-Sigma} as
\beq
\bar{\rho}_\Sigma(\k_F,\bw,\bar{\tau})   &=& \bar{\epsilon}_0 \left(\btau^2 \rho_0(\bw)+\rho_2(\bw)- \alpha (\btau^2 \rho_1(\bw)+ \rho_3(\bw))\right)\label{expandself-energy}
\eeq
where
\beq
\rho_m(\bw) =  e^{-\bw^2} \times \bw^m . \label{rhom}
\eeq

We  write its Hilbert transform as
\beq
\chi(\k_F,\omega,T)&=&\dashint_{-\infty}^\infty \frac{\rho_\Sigma(\k_F,x,T)}{\omega-x} dx\nn \\
&=& \Omega_0 \dashint_{-\infty}^\infty \frac{\bar{\rho}_\Sigma(\k_F, \bar{x},\bar{\tau})}{\bw- \bar{x}} d\bar{x}.
\eeq
By plugging in for $\rho_\Sigma$ we get
\beq
\bar{\chi}(\k_F,\bw,\bar{\tau})\equiv \frac{1}{\Omega_0}\chi(\k_F,\omega,T)=  \bar{\epsilon}_0 \left(\btau^2 \chi_0(\bw)+\chi_2(\bw)- \alpha (\btau^2 \chi_1(\bw)+ \chi_3(\bw))\right) \label{chi},
\eeq
where
\beq
\chi_m(\bw)&=&  \dashint_{-\infty}^\infty \frac{\rho_m(\bar{x} )}{\bw- \bar{x}} d\bar{x} .
\eeq
The evaluation of $\chi_m$ for even $m$ follows from the identity
\beq
\dashint_{-\infty}^\infty \frac{e^{-\alpha y^2}}{x-y} \, dy= 2 \sqrt{\pi} D_F(\sqrt{\alpha} x) \;\;\mbox{ with } \alpha>0,
\eeq
by differentiating under the integral sign with respect to $\alpha$, and where $D_F$ is the Dawson function
\beq
D_F(x)= e^{-x^2} \int_0^x  e^{t^2} \; dt . \label{Dawson-Def}
\eeq
For odd $m$ we use the method of partial fractions to depress the order $m$ by one, and then use the above scheme for order $m-1$. In this way
we find
\beq
\chi_0(x)&=&\sqrt{\pi} 2 D_F(x) \nn \\
\chi_1(x)&=& \sqrt{\pi} [ 2 x D_F(x) -1] \nn \\
\chi_2(x)&=&\sqrt{\pi}  x [2 x D_F(x)-1] \nn \\
\chi_3(x)&= & \sqrt{\pi}[ 2 x^3 D_F(x)-x^2-\half]. 
\eeq
The Dawson function has a series expansion for small $x$
\beq
D_F(x)\sim x- \frac{2}{3} x^3 + O(x^5)
\eeq
We read off $Z$ from this as
\beq
\frac{1}{Z}&=& \alpha_z -  \partial_{\omega} \chi(\k_F,\omega,T)\nn\\
&=&\alpha_z- \bar{\epsilon}_0 \left(\btau^2 \chi'_0(\bw)+\chi'_2(\bw)- \alpha [\btau^2 \chi'_1(\bw)+ \chi'_3(\bw)] \right) \Big\vert_{\bw\to0} \nn\\
&=&\alpha_z+ \sqrt{\pi} \bar{\epsilon}_0 \left(1- 2 \btau^2\right) \label{Z-scaled}
\eeq
where the prime denotes a derivative in the second line.
In dimensionless (i.e. scaled) units with $\bar{\rho}_\Sigma =\Omega_0 \rho_\Sigma$   we write
\beq
\bar{A}(\k_F,\bw,\bar{\tau})= \frac{\frac{\bar{\eta}}{\pi}+\bar{\rho}_{\Sigma}(\k_F,\bw,\bar{\tau})}{[\frac{\bw}{Z(\bw)} -\Delta\bw(\bar{\tau})]^2 + [\bar{\eta}+\pi \bar{\rho}_{\Sigma}(\k_F,\bw,\bar{\tau}) ]^2} \label{spectral-function-scaled}
\eeq
where we have added an  elastic scattering constant $\bar{\eta}$  to  the $\Im \Sigma$, and thereby  a constant  $\frac{\bar{\eta}}{\pi}$ to $\bar{\rho}_\Sigma$ here.
It is important to note that the {\em total self-energy} \disp{rhototal} deduced by inverting the $A$ of \disp{spectral-function-scaled}, will contain an added contribution of $\frac{\eta}{\pi}$  to $\rho_\Sigma$. This is commented on in the Conclusions section \ref{Conclusions}, and made specific in the captions of figures \figdisp{Figure4,Figure5,Figure6,Figure8,Figure10,Figure12} in the paper.

Also note that 
\beq
\Delta\bw(\bar{\tau})= \frac{1}{\Omega_0}(\mu(0)-\mu(T))+ \frac{1}{\Omega_0}(\Sigma_*(\k_F,T)-\Sigma_*(\k_F,0))+ \left(\bar{\chi}(\k_F,0,\bar{\tau}) - \bar{\chi}(\k_F,0,0) \right).\label{delta-omega-unscaled}
\eeq

We note the lowest order $\bw$ expansions
\beq
\bar{\rho}_\Sigma(\k_F,\omega) &\sim &\bar{\epsilon}_0 (\bw^2+ \bar{\tau}^2) (1- \alpha \bw) \nn \\
\bar{\chi}(\bw,\bar{\tau})&\sim& \half \alpha \sqrt{\pi} \bar{\epsilon}_0 (1+ 2 \bar{\tau}^2) - \sqrt{\pi} \bar{\epsilon}_0 (1- 2 \bar{\tau}^2) \bw + {\cal O}(\bw^2)\nn \\ 
\eeq 
It follows that
\beq
\Delta\bw(\bar{\tau})= \frac{1}{\Omega_0}(\mu(0)-\mu(T))+ \frac{1}{\Omega_0}(\Sigma_*(\k_F,T)-\Sigma_*(\k_F,0))+ \alpha \left( \sqrt{\pi} \bar{\epsilon}_0  \bar{\tau}^2 \right)\nn \\ \label{delta-omega-scaled}
\eeq
Therefore the asymmetry parameter $\alpha$ shows up in $\Delta \bw$ linearly.
In some situations it might be reasonable to assume that this term dominates over the others, and if this is prevails then one can expect to extract $\alpha$ from the shift of the spectral peak at $\k_F$ as a function of $T$.

\subsection{  Useful properties of the peaks very close to $\omega=0$}
Here we simplify the above expressions in the neighbourhood of $\omega$=0. 

With 
\beq
Z&=&\frac{1}{\alpha_z+ \sqrt{\pi} \bar{\epsilon}_0 \left(1- 2 \btau^2\right)}\\
\Gamma_0&=&Z \bar{\eta} + Z \bar{\epsilon}_0 \pi \bar{\tau}^2\\
\Gamma_2&=&Z \bar{\epsilon}_0 \pi
\eeq 
we get
\beq
\bar{A}(\k_F,\bw)\sim \frac{Z}{\pi} \frac{\Gamma_0+\Gamma_2 \bw^2}{(\Gamma_0+\Gamma_2 \bw^2)^2+ (\bw- Z \Delta \bw)^2} \label{Fixing-scale}
\eeq
Further  simplifying to  a small  shift $Z \Delta \bw \ll1$ 
\beq
\bar{A}(\k_F,\bw)\sim \frac{Z}{\pi} \frac{\Gamma_0+\Gamma_2 \bw^2}{(\Gamma_0+\Gamma_2 \bw^2)^2+ \bw^2}
\eeq
so that $\{\bar{A}\}_{max}=\bar{A}(\k_F,0)= \frac{Z}{\pi  \Gamma_0}$.
We note that  the scaled width (FWHM) of the spectral peak used in the analysis is given by
\beq
\bar{W}_0=2 \sqrt{-\frac{1}{ 2 \Gamma_2^2}+ \frac{1}{2 \Gamma_2^2}\left( 1+  4 {\Gamma_0^2 \Gamma_2^2}\right)^\half}\label{Width}.
\eeq


\newpage
\vspace{.25 in}
  

 \end{document}